\documentclass[fleqn,twoside]{article}

\usepackage[headings]{espcrc2}

\newskip\humongous \humongous=0pt plus 1000pt minus 100pt
\def\caja{\mathsurround=0pt}
\def\eqalign#1{\,\vcenter{\openup1\jot \caja
       \ialign{\strut \hfil$\displaystyle{##}$&$
        \displaystyle{{}##}$\hfil\crcr#1\crcr}}\,}
\newif\ifdtup

\newcounter{eqnumber}[section]
\renewcommand{\theeqnumber}{\thesection.\arabic{eqnumber}}
\def\equn{\refstepcounter{eqnumber}
\eqno({\rm \theeqnumber})
}

\def\npb#1#2#3{{\rm Nucl. Phys. B}{\bf \ #1}, #3 (#2)}

\def\cqg#1#2#3{{\rm Class. and Quant.\ Grav.} {\bf  #1}, #3 (#2)}

\def\hepth#1{[hep-th/#1]}
\def\hepph#1{[hep-ph/#1]}

\def\eqn#1{eq.~(\ref{#1})}

\newbox\charbox
\newbox\slabox
\def\s#1{{      
        \setbox\charbox=\hbox{$#1$}
        \setbox\slabox=\hbox{$/$}
        \dimen\charbox=\ht\slabox
        \advance\dimen\charbox by -\dp\slabox
        \advance\dimen\charbox by -\ht\charbox
        \advance\dimen\charbox by \dp\charbox
        \divide\dimen\charbox by 2
        \raise-\dimen\charbox\hbox to \wd\charbox{\hss/\hss}
        \llap{$#1$}
}}

\def\spa#1.#2{\left\langle#1\,#2\right\rangle}
\def\spb#1.#2{\left[#1\,#2\right]}
\def\lor#1.#2{\left(#1\,#2\right)}

\catcode`@=11  

\def\la{\langle}
\def\ra{\rangle}
\def\oneloop{{1 \mbox{-} \rm loop}}

\def\lsl{\not{\hbox{\kern-2.3pt $\ell$}}}
\def\ksl{\not{\hbox{\kern-2.3pt $k$}}}

\def\spa#1.#2{\left\langle#1\,#2\right\rangle}
\def\spb#1.#2{\left[#1\,#2\right]}
\def\lor#1.#2{\left(#1\,#2\right)}

\def\sand#1.#2.#3{%
  \left\langle\smash{#1}{\vphantom1}\right|{#2}%
  \left|\smash{#3}{\vphantom1}\right\rangle}
\def\sandp#1.#2.#3{%
  \left\langle\smash{#1}{\vphantom1}^{-}\right|{#2}%
  \left|\smash{#3}{\vphantom1}^{+}\right\rangle}
\def\sandpp#1.#2.#3{%
  \left\langle\smash{#1}{\vphantom1}^{+}\right|{#2}%
  \left|\smash{#3}{\vphantom1}^{+}\right\rangle}
\def\sandmm#1.#2.#3{%
  \left\langle\smash{#1}{\vphantom1}^{-}\right|{#2}%
  \left|\smash{#3}{\vphantom1}^{-}\right\rangle}
\def\sandpm#1.#2.#3{%
  \left\langle\smash{#1}{\vphantom1}^{+}\right|{#2}%
  \left|\smash{#3}{\vphantom1}^{-}\right\rangle}
\def\sandmp#1.#2.#3{%
  \left\langle\smash{#1}{\vphantom1}^{-}\right|{#2}%
  \left|\smash{#3}{\vphantom1}^{+}\right\rangle}

\def\BR#1#2{\la#1|{P_{abc}}|#2\ra}

\def\tree{{\rm tree}}

\def\NeqEight{{\cal N} = 8}
\def\NeqFour{{\cal N} = 4}
\def\NeqOne{{\cal N} = 1}

\def\Fact{{\cal F}}

\title{Recursive Approach to One-loop QCD Matrix Elements}

\author{Zvi Bern\address[UCLA]{Department of Physics,
University of California
at Los Angeles,},
N.~E.~J.~Bjerrum-Bohr\address[UWS]{Department of Physics, 
University of Wales Swansea},
David~C.~Dunbar\addressmark[UWS]
\address[Presenter]{Presented by D. Dunbar at RADCOR 2005}
and
Harald~Ita\addressmark[UWS] 
}

\begin{document}

\maketitle

\section{Introduction}

\vskip -0.2 truecm 
Recently, a ``weak-weak'' duality, between ${{\cal N}=4}$ super
Yang-Mills and a topological string theory propagating in twistor
space, has been proposed~\cite{Witten:2003nn} implying an identical
perturbative $S$-matrix for the two theories. 
The existence of a duality between the
two theories implies a surprising structure within the $S$-matrix
of gauge theory.  This has inspired
considerable progress in computing scattering amplitudes.

The generalisation of these ideas combined with ideas from the
unitarity method~\cite{BDDKa,BDDKb} has led to new ideas in
computing one-loop gluon scattering
amplitudes~\cite{BDKrecursionA,BDKboot,Bern:2005hh} in theories with less than
maximal or no supersymmetry such as massless QCD.  In this talk we
discuss and review this work with particular reference to the results
for one-loop QCD amplitudes~\cite{BDKrecursionA,Bern:2005hh}.
The particular approach that we describe is recursive and our aim is
to establish recursion relations where an $n$-point one-loop amplitude is
obtained from expressions for lower-point amplitudes, bypassing 
the need for performing any loop integrations.  As yet, 
this approach only works in cases where certain criteria 
on the unitarity cuts are satisfied.  But in the cases where the
criteria are satisfied, it is a particularly effective.

The duality is most obvious if we express the amplitude in terms of
fermionic ``twistor'' variables.  We can achieve this by replacing
everywhere the massless momentum $p_{a\dot a}$ by $
\lambda_a\bar\lambda_{\dot a}$ where $p_{a\dot a}=(\sigma^\mu)_{a\dot
a} p_\mu$.  The external polarisation vectors can also be defined in
terms of spinor variables~\cite{SpinorHelicity} using the
spinor-helicity notation.
%
%

This talk is primarily about loop calculations, however, there are 
two twistor inspired techniques for computing tree amplitudes 
which we wish to discuss.  First there is the MHV-vertex 
construction by Cachazo, Svrcek and Witten (CSW)~\cite{CSW} and 
secondly there is the recursion relations by Britto, Cachazo, 
Feng and Witten (BCFW)~\cite{Britto:2004ap}. 

In the MHV vertex approach, amplitudes are obtained by sewing together
``MHV vertices''. A $n$-point MHV vertex has exactly two gluons of
negative helicity and all remaining helicities positive. 
Amplitudes with more negative helicities, for example, next-to-MHV or 
`NMHV'' amplitudes, 
are in the CSW formalism constructible from products of 
MHV vertices. The forms of these vertices are those of the Parke-Taylor 
amplitudes~\cite{ParkeTaylor} where a specific off-shell continuation 
is employed for the internal particle lines. 
The CSW amplitude construction is explicitly asymmetric in gluon helicity. 

This formalism is a remarkable rewriting of perturbation theory.
It has been extended to
a variety of cases beyond that of gluon
scattering~\cite{MHVextensions}. The MHV amplitude has been shown to
extend to one-loop amplitudes within supersymmetric
theories~\cite{MHV1loop} although application of these rules still
requires integration and an extension to non-supersymmetric theories
proves more difficult.

The BCFW recursion relations~\cite{Britto:2004ap} rely on the analytic 
structure of the amplitude after it has been continued to a function in 
the complex plane $A(z)$. This continuation is a shift in the
(spinorial) momentum of two chosen legs,
$$
p^1_{a\dot a} \longrightarrow 
p^1_{a\dot a} + z \lambda^1_a\bar\lambda^2_{\dot a} 
\;,\;\;  
p^2_{a\dot a} \longrightarrow 
p^2_{a\dot a} - z \lambda^1_a\bar\lambda^2_{\dot a}\,. 
\equn
$$
By integrating $A(z)/z$ over a contour at infinity and assuming 
$A(z) \longrightarrow 0$, 
$A(0)$ can be determined from the remaining poles of the function 
$A(z)/z$ at $z=z_i\neq 0$. 
The poles of this function at $z_i$ are given by the factorisations 
of the amplitude $A(z)$ which occur where some intermediate momenta 
$P(z)$ becomes on shell, {\it i.e.,} 
$P(z)^2=0$ for some intermediate $P(z)$. The residue 
is given by the product of two tree amplitudes and 
we thus obtain the recursion relation which gives the $n$-point amplitude 
as a sum over (shifted) lower point functions
$$
A(0)=  \sum _{i,h} \hat A^h_k(z_i)  \times { i \over P^2}  \times \hat A^{-h}_{n-k+1}(z_i)\,.
\equn\label{TreeRecursion}
$$
The above summation only includes trees where 
the two shifted legs $1$ and $2$
are on opposite sides of the poles. 
The tree amplitudes are evaluated at
the value of $z$ such that the shifted pole term vanishes, 
{\it i.e.} $P(z)^2=0$.  
The analytic structure of the amplitude is the key ingredient in this process. 
The techniques also extend to many situations  
and 
in fact the correctness of the MHV construction can be derived from this 
approach~\cite{Britto:2004ap,Kasper,Bjerrum-Bohr:2005jr}.
The BCFW recursion relations differ from the well established Berends-Giele
recursion relations~\cite{BerendsGiele} in that they are on-shell. 

In this talk we will be interested in extending the above technique 
to one-loop amplitudes. We will pursue the possibility of recursive techniques 
which avoid integration of loop momenta.

%
\section{Supersymmetric Decomposition of QCD Amplitudes} 

In general, we examine colour-decomposed amplitudes. 
Let $A_{n}^{[J]}$ denote the leading in colour 
partial amplitude for gluon scattering due to an
(adjoint) particle of spin $J$\, in the loop. 
The three choices we are interested in are gluons ($J=1$), adjoint 
fermions ($J=1/2$) and adjoint scalars ($J=0$). It is considerably
easier to calculate the contributions due to supersymmetric matter
multiplets together with the complex scalar. 
The three types of supersymmetric multiplets are the $\NeqFour$ 
multiplet and the $\NeqOne$ vector and matter multiplets. 
These contributions are related to the $A_{n}^{[J]}$ by~\cite{FiveGluon}
$$
\eqalign{
A_{n}^{\,\NeqFour} &\;\equiv\;
A_{n}^{[1]}\; +\; 4A_{n}^{[1/2]}\;+\;3 A_{n}^{[0]}\,,
\cr
A_{n}^{\,\NeqOne\; {\rm vector}} &\;\equiv\; A_{n}^{[1]}\
\;+\;A_{n}^{[1/2]} \,,
\cr
 A_{n}^{\,\NeqOne\; {\rm chiral}} &\;\equiv\;
A_{n}^{[1/2]}\; +\; A_{n}^{[0]} \,.
\cr}
\equn
$$
These relations can be inverted to obtain the amplitudes for QCD via
$$
\eqalign{
A_{n}^{[1]} &\; = \; A_{n}^{\,\NeqFour}-4A_{n}^{\,\NeqOne\; {\rm chiral}}\;+\;A_{n}^{[0]}\,,
\cr
A_{n}^{[1/2]} &\; = \; A_{n}^{\,\NeqOne\; {\rm chiral}}\;-\;A_{n}^{[0]}\,.
\cr}
\equn
$$ 
The contribution from massless quark scattering can be obtained from
these.  When computing amplitudes in supersymmetric theories we are
calculating well defined pieces of QCD amplitudes -- although the
procedure is incomplete unless we can obtain the non-supersymmetric
$A_{n}^{[0]}$.


\section{$\NeqFour$ Contribution}
In the supersymmetric amplitudes there are generically cancellations 
between the bosons and fermions in the loop.  For $\NeqFour$ SYM 
these cancellations lead to considerable simplifications in the 
loop momentum integrals. This is
manifest in the ``string-based approach'' for computing loop
amplitudes~\cite{StringBased}.  As a result of these simplifications,
$\NeqFour$ one-loop amplitudes can be expressed simply as a sum of
scalar box-integral functions, $I_4^i$,~\cite{BDDKa}.
$$
A^{\NeqFour}  = \sum_{i}  c_i I_4^i \,,
\equn\label{generalform2}
$$
and the computation of one-loop $\NeqFour$ amplitudes is then  
a matter of determining the rational coefficients $c_i$.

The box-coefficients are ``cut-constructible''~\cite{BDDKa}. That is
they may be determined by an analysis of the cuts.  This allows a
variety of techniques to be used in evaluating these.  Originally an
analysis of unitary cuts was used to determine the coefficients
firstly for the MHV case~\cite{BDDKa} and secondly for the remaining
six-point amplitudes~\cite{BDDKb}.  The unitarity method 
when combined with twistor inspired ideas, 
has led to rapid development of new computational methods over
the past year~\cite{BeDeDiKo,Britto:2004nj,Roiban:2004ix,BrittoUnitarity,BDKn,Bidder:2005in}.
The box-coefficients often display non-trivial geometric structure in
twistor space such as collinearity or 
coplanarity~\cite{Witten:2003nn,BDKn,Britto:2004tx,BBDP}.  
Related to this is the conjecture that $\NeqEight$ supergravity amplitudes
are also only composed of box integral functions~\cite{GravityBoxes}, which may
lead to a reconsideration of the ultraviolet infinity structure of this theory~\cite{BDDPR}. 

\section{$\NeqOne$ Contribution} 

We shall keep this section brief by necessity although much
interesting progress has been made for the remaining 
supersymmetric contributions.  
$\NeqOne$ amplitudes are also cut constructible~\cite{BDDKb}
although in this case the integral functions are more complicated 
involving additionally scalar triangle and bubble functions. 
$$
A^{\NeqOne}= \sum_i c_i I^i_4 
+\sum_i d_i I^i_3 
+\sum_i e_i I^i_2\,. 
\equn
$$
The five point amplitudes in this case have been known for some time.  
The recent progress has seen the computation of the six-point 
amplitudes including the NMHV cases~\cite{Bidder:2004tx,BBDP,BrittoSQCD,Bern:2005hh}.
 
\section{The non-supersymmetric parts of QCD Amplitudes}
From the supersymmetric decomposition the calculation of a gluon
scattering amplitude may be reduced to that for a scalar circulating
in the loop. This amplitude is not cut-constructible but can be expanded
$$
A^{[0]}= \sum_i c_i I^i_4 
+\sum_i d_i I^i_3 
+\sum_i e_i I^i_2 
+R\,,
\equn
$$
where $R$ denotes the rational terms which are not cut-constructible
unless one determines the cuts beyond the leading orders in the dimensional regularisation parameter $\epsilon\equiv(4-D)/2$~\cite{BDDKmassive}.  

Loop amplitudes contain logarithmic (and dilogarithmic) terms which
would contain cuts in the complex plane when shifted. Thus, in general,
the entire
amplitude cannot be described by a recursion relation of the type in 
\eqn{TreeRecursion}. However,
there are two places where this type of analytic recursion relation may be
used.

{\bf A} \ The rational terms $R$~\cite{BDKrecursionA}
$$
R \equiv (  A^{[0]}- \sum_i c_i I^i_4 +\sum_i d_i I^i_3  +\sum_i e_i I^i_2  )\,.
\equn
$$

{\bf B} \ The rational coefficients of the integral functions $c_i$, $d_i$ and 
$e_i$~\cite{Bern:2005hh}.   

 \def\inlimit^#1{\buildrel#1\over\llongrightarrow}
\def\llongrightarrow{%
\relbar\mskip-0.5mu\joinrel\mskip-0.5mu\relbar    
\mskip-0.5mu\joinrel\longrightarrow}

In this talk we focus on the second case.
In both cases, in order to apply recursion relations one has to understand 
the pole structure of the amplitude.  
The full amplitude obeys a factorisation given by~\cite{BernChalmers}
$$
\eqalign{
A_{n}^{\oneloop}\
& \inlimit^{P_{i,i+m-1}^2\rightarrow 0} \cr & \hspace{-0.5cm}
\sum_{}  \Biggl[
 A_{m+1}^{\oneloop} \,
            {i \over P_{i,i+m-1}^2} \,
   A_{n-m+1}^{\tree} \cr
& \hskip2.8cm \null\hspace{-0.9cm}
 + A_{m+1}^{\tree} \, {i\over P_{i,i+m-1}^2} \,
   A_{n-m+1}^{\oneloop}
\label{LoopFact} \cr
& \hskip-0.1cm \null
 + A_{m+1}^{\tree} \, {i\over P_{i,i+m-1}^2} \,
   A_{n-m+1}^{\tree} \,
      \Fact_n \Biggr] \,,
\cr}\equn
\label{factorisation}
$$
which specifies the singularities of the rational terms and
the rational coefficients.  

In addition to physical singularities, pieces of amplitudes also
contain {\it spurious singularities}.  A spurious singularity is a
singularity which does not appear in the full amplitude but which
appears in the various components. Typical examples are co-planar
singularities such as
$$
{ 1 \over \BR23  }
\equn
$$
which vanishes when $P_{abc}=\alpha k_2 +\beta k_3$. 
Such singularities are common in the coefficients of integral functions.
On these singularities, the integral functions are not independent but 
combine to cancel. For example, for six-point kinematics, the product $\la 2 |
P_{234} | 5 \ra$ vanishes when $t_{234}t_{612}-s_{34}s_{61}=0$.  At
this point the functions $\ln(s_{34}/t_{234})$ and
$\ln(s_{61}/t_{612})$ are no longer independent and the combination
$$
{ a_1 \over\la 2 | P_{234}  | 5 \ra } \ln(s_{34}/t_{234}) +
{a_2\over \la 2 | P_{234}  | 5 \ra }\ln(s_{61}/t_{612})\,,
\equn
$$ is non-singular provided $a_1=a_2$ at the singularity.
Such spurious singularities are best avoided by a careful choice of shift
for a specific integral function. 
 
Spurious singularities are also related to 
the choice of
basis functions. For example expressions such as $\ln(r)/(1-r)^3$
typically appear in the cut-constructible part of the amplitude where $r$ is
same ratio of kinematic variables.
These expressions are singular at $r=1$ which does not normally correspond to 
a physical pole. These spurious singularities
cancel between these terms and the rational terms. 
If we instead choose an improved basis function $L_2(r)=(
\ln(r)+(r-r^{-1}) )/(1-r)^3$ which is finite as $r\longrightarrow 1$
then both the cut-constructible and rational terms will be free of
this spurious singularity.

Assuming that the spurious denominators do not pick up
a $z$ dependence -- in ref.~\cite{Bern:2005hh}
we describe simple criteria based on the unitarity cuts for ensuring this  --
we obtain a recursion relation for the coefficients
analogous to that for tree amplitudes,
$$
c_n(0) \; = \; \sum_{\alpha,h}  {A^h_{n-m_\alpha+1}(z_\alpha) \,
 {i\over P^2_{\alpha}}\, c^{-h}_{m_\alpha+1}(z_\alpha)} \,,
\equn\label{CoeffRecur}
$$
where $A^h_{n-m_\alpha +1}(z_\alpha)$ and $c^h_{n-m_\alpha+1}(z_\alpha)$ are
shifted tree amplitudes and coefficients evaluated at the residue
value $z_\alpha$\,, $h$ denotes the helicity of the intermediate state
corresponding to the propagator term $i/P^2_{\alpha}$\,. In this
expression one should only sum over the limited set of poles that
can appear in the integral coefficients.
This has successfully been
applied~\cite{Bern:2005hh} to determine the integral coefficients,
$\hat d_{n,r}$, $\hat g_{n,r}$ and $\hat h_{n,r}$ in 
the amplitude
\def\Lzz{L_2}
{\small
$$
\eqalign{
&A_n^{[0]}(1^-,2^-,3^-,4^+,5^+,\cdots, n^+)\; = \;
 \cr
&\frac{1}{3}\,A_{n}^{\,\NeqOne\ {\rm chiral}}(1^-,2^-,3^-,4^+,5^+,\cdots, n^+) \cr
&
-{i \over 3}
\sum_{r=4}^{n-1}\, \hat d_{n,r}\,
{    \Lzz [ t_{3,r} / t_{2,r} ] \over t_{2,r}^3  }
-{i \over 3}
\sum_{r=4}^{n-2}\, \hat g_{n,r}\,
{    \Lzz [ t_{2,r} / t_{2,r+1} ] \over t_{2,r+1}^3  }
\cr
&-{i \over 3}
\sum_{r=4}^{n-2}\ \hat h_{n,r}
{    \Lzz [ t_{3,r} / t_{3,r+1} ] \over t_{3,r+1}^3  }
+ \hbox{rational}\,, \cr
}
\equn
$$
}together with the extension to the ``split helicity'' configuration
$A(1^-,\cdots,r^-,r+1^+,\cdots ,n^+)$.  The rational terms should also
be obtainable using recursion, following the methods of
ref.~\cite{BDKrecursionA,LanceQMUL}.  This would achieve our goal of
avoiding all loop integrations to obtain these amplitudes.

\section{Summary of Six-gluon Amplitude} 

It is pertinent to ask how the new techniques are contributing to new
calculations with QCD. At one loop the four and five gluon amplitudes
are known~\cite{EllisSexton,FiveGluon} however the six-gluon is not
yet completely calculated analytically.  The above table summarises the
``state of play`` in this calculation.  (There has also been some very 
recent progress with semi-numerical 
methods~\cite{EGZ}, providing a check on the above calculations.)
 The amplitude is split
into the two supersymmetric contributions plus the scalar piece. The
scalar is further subdivided into the cut constructible integral
functions $S_C$ together with the rational pieces $S_R$. 
In the past year, much progress has occurred, although much more remains
to be done, to apply these ideas to problems in collider physics.

\begin{table}
\caption{The Status of the Six-Gluon Amplitude}
\begin{tabular}{|l|l|l|l|l|} \hline
   & $\NeqFour$    & $\NeqOne$     
& $ S_C $  &   $ S_R $
\\[2pt]
\hline
 {\small  $A(--++++)$ }  &  \cite{BDDKa}     &
\cite{BDDKb}    & \cite{BDDKb}    &  \cite{BDKboot} 
\\[2pt]
\hline
 {\small $A(-+-+++) $ }  & \cite{BDDKa}     &
\cite{BDDKb} & \cite{Bedford:2004nh}  &  
\\[2pt]
\hline
  {\small $A(-++-++)$ }  & \cite{BDDKa}     &
\cite{BDDKb}   & \cite{Bedford:2004nh} &   
\\[2pt]
\hline
  {\small $A(---+++)$ }  &  \cite{BDDKb}   &
\cite{Bidder:2004tx}  & \cite{Bern:2005hh}  & \cite{LanceQMUL}  
\\[2pt]
\hline
  {\small $A(--+-++)$}  &  \cite{BDDKb}    &
\cite{BrittoSQCD,BBDP}    & \cite{BFM} &  
\\[2pt]
\hline
 {\small $A(-+-+-+)$}  &   \cite{BDDKb}   &
\cite{BrittoSQCD,BBDP}    & \cite{BFM}  &  
\\[2pt]
\hline
\end{tabular}
\end{table}
\section{Conclusions}

The past two years have seen significant progress in the computation
of loop amplitudes in gauge theories. Although, many of these techniques have
arisen in the context of supersymmetric theories, the process of
applying them to theories such as QCD is underway, with the first
concrete results for one-loop amplitudes in QCD with six or more external 
particles now appearing.


\begin{thebibliography}{99}


\bibitem{Witten:2003nn}
E.~Witten,
Commun.\ Math.\ Phys.\  {\bf 252}, 189 (2004)
[hep-th/0312171].


\bibitem{BDDKa}
Z. Bern, L.J. Dixon, D.C. Dunbar and D.A. Kosower,
\npb{425}{1994}{217}, \hepph{9403226}.

\bibitem{BDDKb}
Z. Bern, L.J. Dixon, D.C. Dunbar and D.A. Kosower,
\npb{435}{1995}{59}, \hepph{9409265}.


\bibitem{BDKrecursionA}
Z.~Bern, L.J.~Dixon and D.A.~Kosower,
Phys.\ Rev.\ D {\bf 71}, 105013 (2005)
[hep-th/0501240];
%
Phys.\ Rev.\ D {\bf 72}, 125003 (2005)
[hep-ph/0505055];
D.~Forde and D.~A.~Kosower,
hep-ph/0509358.

\bibitem{BDKboot}
 Z.~Bern, L.~J.~Dixon and D.~A.~Kosower,
hep-ph/0507005.


\bibitem{Bern:2005hh}
 Z.~Bern, N.~E.~J.~Bjerrum-Bohr, D.~C.~Dunbar and H.~Ita,
  JHEP {\bf 0511}, 027 (2005)
  [hep-ph/0507019].



\bibitem{SpinorHelicity}
Z.~Xu, D.~H.~Zhang and L.~Chang,
Nucl.\ Phys.\ B {\bf 291}, 392 (1987).


\bibitem{CSW}
F.~Cachazo, P.~Svrcek and E.~Witten,
JHEP {\bf 0409}, 006 (2004) [hep-th/0403047].


\bibitem{Britto:2004ap}
R.Britto, F.Cachazo and B.Feng,
hep-th/0412308; \\
R.~Britto, F.~Cachazo, B.~Feng and E.~Witten,
hep-th/0501052.



\bibitem{ParkeTaylor}
S.J. Parke and T.R. Taylor,
Phys.\ Rev.\ Lett.\ 56:2459
(1986).
%

\bibitem{MHVextensions}
G.~Georgiou and V.~V.~Khoze,
JHEP {\bf 0405}, 070 (2004)
[hep-th/0404072];\\
%
%
J.~B.~Wu and C.~J.~Zhu,
JHEP {\bf 0409}, 063 (2004) [hep-th/0406146];\\
%
%
X.~Su and J.~B.~Wu,
hep-th/0409228;\\
%
L.~J.~Dixon, E.~W.~N.~Glover and V.~V.~Khoze,
JHEP {\bf 0412}, 015 (2004)
[hep-th/0411092];\\
Z.~Bern, D.~Forde, D.~A.~Kosower and P.~Mastrolia,
hep-ph/0412167;\\
S.~D.~Badger, E.~W.~N.~Glover and V.~V.~Khoze,
hep-th/0412275.


\bibitem{MHV1loop}
A.~Brandhuber, B.~Spence and G.~Travaglini,
 Nucl.\ Phys.\ B {\bf 706}, 150 (2005) [hep-th/0407214];\\
C.~Quigley and M.~Rozali,
 JHEP {\bf 0501}, 053 (2005) [hep-th/0410278];\\
J.~Bedford, A.~Brandhuber, B.~Spence and G.~Travaglini,
 Nucl.\ Phys.\ B {\bf 706}, 100 (2005)
 [hep-th/0410280].



\bibitem{Kasper}
K. Risager,
hep-th/0508208


\bibitem{Bjerrum-Bohr:2005jr}
  N.E.J.~Bjerrum-Bohr, D.C.Dunbar, H.Ita, W.B.Perkins and K.Risager,
  hep-th/0509016.



\bibitem{BerendsGiele}
F.~A.~Berends and W.~T.~Giele,
Nucl.\ Phys.\ B {\bf 306}, 759 (1988).


\bibitem{FiveGluon}Z. Bern, L. Dixon and D.A.\ Kosower, Phys.\ Rev.\ Lett.\
70:2677 (1993).


\bibitem{StringBased}
Z.~Bern and D.~A.~Kosower,
Phys.\ Rev.\ Lett.\  {\bf 66}, 1669 (1991).
Nucl.\ Phys.\ B {\bf 379}, 451 (1992);\\
Z. Bern, D.C. Dunbar and T. Shimada, 
  Phys.\ Lett.\ B {\bf 312}, 277, (1993)
[hep-th/9307001];\\
Z.~Bern and D.~C.~Dunbar,
Nucl.\ Phys.\ B {\bf 379}, 562 (1992);\\
D.C. Dunbar and P.S. Norridge,
\npb{433}{1995}{181} [hep-th/9408014].



\bibitem{BeDeDiKo}
Z.~Bern, V.~Del Duca, L.J.~Dixon and D.~A.~Kosower,
 Phys.\ Rev.\ D {\bf 71}, 045006 (2005)
[hep-th/0410224].


\bibitem{Britto:2004nj}
R.Britto, F.Cachazo and B.Feng,
hep-th/0410179.

\bibitem{Roiban:2004ix}
R.~Roiban, M.~Spradlin and A.~Volovich,
hep-th/0412265.


\bibitem{BrittoUnitarity}
R.~Britto, F.~Cachazo and B.~Feng,
hep-th/0412103.



\bibitem{BDKn}
Z.~Bern, L.~J.~Dixon and D.~A.~Kosower,
hep-th/0412210.


\bibitem{Bidder:2005in}
  S.~J.~Bidder, D.~C.~Dunbar and W.~B.~Perkins,
  hep-th/0505249.


\bibitem{Britto:2004tx}
R.~Britto, F.~Cachazo and B.~Feng,
hep-th/0411107.






\bibitem{BBDP}
S.~J.~Bidder, N.~E.~J.~Bjerrum-Bohr, D.~C.~Dunbar and
W.~B.~Perkins,
hep-th/0412023;
hep-th/0502028.

\bibitem{GravityBoxes}
Z.~Bern, N.~E.~J.~Bjerrum-Bohr and D.~C.~Dunbar,
hep-th/0501137;\\
  N.~E.~J.~Bjerrum-Bohr, D.~C.~Dunbar and H.~Ita,
  hep-th/0503102.

\bibitem{BDDPR}
Z. Bern, L.J. Dixon, D.C. Dunbar, M.\ Perelstein and J.S.\ Rozowsky,
\npb{530}{1998}{401} \hepth{9802162};
\cqg{17}{2000}{979}
\hepth{9911194};\\
P.~S.~Howe and K.~S.~Stelle,
 Phys.\ Lett.\ B {\bf 554}, 190 (2003)
 [hep-th/0211279].




\bibitem{Bidder:2004tx}
  S.~J.~Bidder, N.~E.~J.~Bjerrum-Bohr, L.~J.~Dixon and D.~C.~Dunbar,
  Phys.\ Lett.\ B {\bf 606}, 189 (2005)
[hep-th/0410296].


\bibitem{BrittoSQCD} 
R.~Britto, E.~Buchbinder, F.~Cachazo and B.~Feng,
hep-ph/0503132.


\bibitem{BDDKmassive}
Z. Bern, L. Dixon, D.C. Dunbar and D.A. Kosower,
Phys.\ Lett.\ B {\bf 394}, 105 (1997)
[hep-th/9611127].



\bibitem{BernChalmers}
Z.~Bern and G.~Chalmers,
Nucl.\ Phys.\ B {\bf 447}, 465 (1995)
[hep-ph/9503236].


\bibitem{LanceQMUL} 
Lance Dixon, presentation at ``From Twistors to Amplitudes'' at QMUL,
3-5 Nov. 2005


\bibitem{EllisSexton}
R.~K.~Ellis and J.~C.~Sexton,
  Nucl.\ Phys.\ B {\bf 269}, 445 (1986).



\bibitem{Bedford:2004nh}
  J.~Bedford, A.~Brandhuber, B.~J.~Spence and G.~Travaglini,
  Nucl.\ Phys.\ B {\bf 712}, 59 (2005)
  [hep-th/0412108].


\bibitem{BFM}
R.~Britto, B.~Feng and P.~Mastrolia,
  arXiv:hep-ph/0602178.


\bibitem{EGZ}
R.~K.~Ellis, W.~T.~Giele and G.~Zanderighi,
hep-ph/0602185.























\end{thebibliography}
\end{document}